\documentclass[letterpaper, 10 pt, journal, twoside]{IEEEtran}

\usepackage{graphicx} 
\usepackage{amsmath} 
\usepackage{amssymb}  
\usepackage{amsfonts}
\usepackage{algorithmic}
\usepackage{algorithm}
\usepackage{bm}
\usepackage{color}
\usepackage{xcolor}

\begin{document}
\markboth{IEEE Robotics and Automation Letters. Preprint Version. Accepted February, 2025}
{Yokoe \MakeLowercase{\textit{et al.}}: Intuitive Hand Positional Guidance Using McKibben-Based Surface Tactile Sensations to Shoulder and Elbow} 

\title{
	Intuitive Hand Positional Guidance Using McKibben-Based Surface Tactile Sensations to Shoulder and Elbow 
}

\author{Kenta Yokoe$^{1, 2}$, Yuki Funabora$^{3}$, and Tadayoshi Aoyama$^{4, 5}$\\
\thanks{Manuscript received: October 2, 2024; Revised December 26, 2024; Accepted February 1, 2025.}
\thanks{This paper was recommended for publication by Editor Jee-Hwan Ryu upon evaluation of the Associate Editor and Reviewers' comments.} 
	\thanks{This work was supported in part by JST [Moonshot R\&D] Grant Number JPMJMS2214-08 (development for base system), in part by JSPS KAKENHI Grant Numbers JP23K24886 (experiment),  JP23KJ1116 (experiment), JP23K17450 (development for fabric actuators).}
	\thanks{The protocol of this study was approved by the Ethics Committee of the Graduate School of Engineering, Nagoya University (Approval Numbers 23-11 and 24-6).}
	\thanks{$^{1}$K. Yokoe is with the Department of Micro-Nano Mechanical Science and Engineering, Nagoya University, Furo-cho, Chikusa-ku, Nagoya, Aichi 464-8603, Japan.
		{\tt\small yokoe@nagoya-u.jp}}%
	\thanks{$^{2}$K. Yokoe is a Research Fellow of the Japan Society for the Promotion of Science, Tokyo, 102-0083, Japan. }%
	\thanks{$^{3}$Y. Funabora is with the Department of Information and Communication Engineering, Nagoya University
		, Furo-cho, Chikusa-ku, Nagoya, Aichi 464-8603, Japan.
		{\tt\small funabora@nagoya-u.jp}}%
	\thanks{$^{4}$T. Aoyama is with the Department of Mechanical Systems Engineering, Nagoya University, Furo-cho, Chikusa-ku, Nagoya, Aichi 464-8603, Japan.
		{\tt\small tadayoshi.aoyama@mae.nagoya-u.ac.jp}}
	\thanks{$^{5}$T. Aoyama is with the Center for One Medicine Innovative Translational Research, Gifu University, Yanagido 1-1, Gifu, Gifu 501-1193, Japan.}
	\thanks{Digital Object Identifier (DOI): 10.1109/LRA.2025.3540579}
	\thanks{This is the accepted version of an article published in IEEE Robotics and Automation Letters, vol.~10, no.~4, pp.~3254--3261, 2025, DOI: 10.1109/LRA.2025.3540579. \copyright~2025 The Authors. Open Access under CC BY-NC-ND 4.0 (https://creativecommons.org/licenses/by-nc-nd/4.0/).}
}

	\maketitle

	\begin{abstract}
		Hand positional guidance with intuitive perception is crucial for enhancing user interaction and task performance in immersive environments. 
		However, conventional hand positional guidance methods, relying on tactile sensations, lack intuitiveness. 
		Consequently, users require instruction on the relationship between the tactile sensation and target position of the guidance before using these methods. Additionally, the user needs training to become familiar with tactile sensations. 
		This study presents a hand positional guidance system with intuitive perception that leverages McKibben-based surface tactile sensations directed to the shoulder and elbow. 
		We developed a wearable fabric actuator that provides McKibben-based surface tactile sensations to induce six specific movements: elbow flexion, extension, shoulder abduction, adduction, horizontal abduction, and horizontal adduction. 
		The effectiveness of the actuator was experimentally validated, demonstrating its high accuracy in intuitively inducing six movements. 
		An algorithm based on the equilibrium point hypothesis and Weber--Fechner law was implemented to regulate the intensity of the tactile sensations for hand positional guidance. 
		Furthermore, the accuracy and speed of the proposed system were compared with that of conventional guidance methods utilizing synthesized speech and vibrotactile guidance.  
	\end{abstract}
	
	
	\section{Introduction}
	\begin{figure*}[!t]
		\centering
		\includegraphics[keepaspectratio=true,width=.8\linewidth]{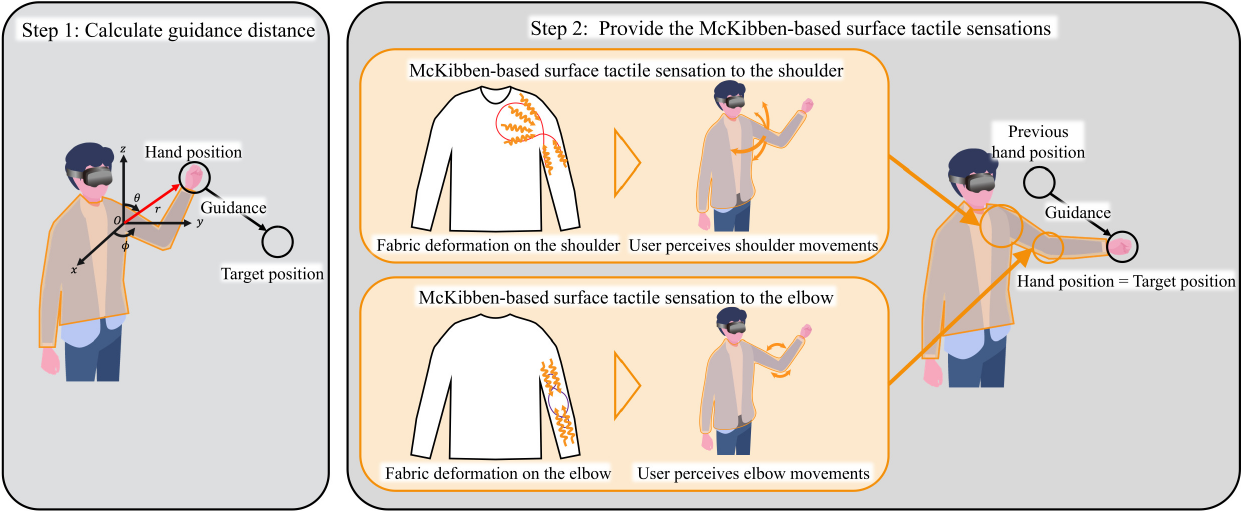}
		\caption{Schematic of the proposed hand positional guidance system.}
		\label{fig:concept} 
	\end{figure*}
	\IEEEPARstart{O}{ver} the past few decades, many studies have been conducted on immersive environments \cite{rubio-tamayo2017Immersive}, with particular emphasis 
	on the teleoperation of robots and other devices in such environments \cite{tavakkoli2020Immersive}. 
	These teleoperation systems facilitate intuitive and seamless operations in environments where users are not physically present, such as microenvironments \cite{yokoe2022immersive}.
	The sense of presence (i.e., the sense of being there) in such immersive environments affects a user's work efficiency \cite{cummings2016How}.
	Immersive environment systems frequently integrate hand tracking with vision or other devices to allow users to interact with virtual objects using their hands to improve the sense of presence \cite{spittle2023Review}.
	Therefore, presenting information pertaining to hand movements is imperative to ensure smooth operation within immersive environments.
	
	Although several hand positional guidance systems offer visual cues \cite{saunders2004Visual},  the human capacity for processing visual information is limited \cite{cohen2014Processing}. 
	Furthermore, hand positional guidance should not interfere with hand interactions or visual hand tracking.
	Therefore, this study proposes a hand positional guidance system that uses tactile sensations in areas other than the hands.
	
	Fig.~\ref{fig:concept} shows a schematic of the proposed hand positional guidance system. 
	First, we developed a wearable fabric actuator that provides McKibben-based surface tactile sensations that induce four specific shoulder movements: abduction, adduction, horizontal abduction, and horizontal adduction. The experimental evaluation of the actuator's performance demonstrated that the actuator was highly accurate in inducing shoulder movements without bias. 
	Next, we developed a hand positional guidance system by integrating the aforementioned actuator with another actuator mentioned in our previous work, providing McKibben-based surface tactile sensations for elbow movements \cite{yokoe2024Elbow}.
	Thus, the proposed system induces six movements: four shoulder movements and two elbow movements.
	Furthermore, we implemented an algorithm that models the hand position using a spherical coordinate system and adjusts the intensity of the tactile sensation based on psychophysics, allowing users to perceive sensations and adjust their hand positions intuitively.
	Finally, the proposed system was experimentally evaluated for speed and accuracy of hand positional guidance through user participation. 
	
	\section{Related Work}
	\label{sec:related}
	Various tactile presentation methods have been proposed for hand positional guidance, including methods that utilize direct tactile sensations to the hands \cite{gil2022Enabling,pascher2023HaptiX}.
	However, for the user to perceive the tactile sensation intuitively,  the device must be attached directly to the hand, which may interfere with hand tracking.
	Therefore, several methods exist for providing guidance through tactile presentation to the wrists \cite{salazarluces2018PhantomSensation,salazar2018PathFollowing} and forearms, \cite{aggravi2016Haptic,elsayed2023Tactileb} that can coexist with hand tracking.
	Based on the leading joint hypothesis \cite{dounskaia2010Control}, these methods typically use the wrist as the leading joint, with the shoulder and elbow functioning as subordinate joints. However, users must learn to interpret the tactile feedback before use.
	
	McKibben-based surface tactile sensation uses McKibben artificial muscles to deform the cloth and create a tactile sensation \cite{yokoe2024Elbow,peng2023FunabotSuit}. 
	The user can perceive this tactile sensation more intuitively compared to conventional tactile sensations such as vibrotactile sensations.
	Therefore, users need no instructions on the relationship between the type of tactile sensation and the target position. In addition, users need no training to get accustomed to tactile sensation. 
	In this study, we applied McKibben-based surface tactile sensation to hand guidance, using the shoulder and elbow as leading joints, with the wrist and hand as subordinate joints. This approach offers more intuitive guidance than conventional wrist-based systems.
	
	\section{Wearable Fabric Actuator Providing McKibben-based Surface Tactile Sensations that Induce Shoulder Movement}
	\label{sec:shoulder}
	\subsection{Configuration of Fabric Actuator for Inducing Shoulder Movements}
	\label{subsec:config}
	McKibben-based surface tactile sensations stimulate the body areas that control specific movements, inducing the wearer to perform these movements \cite{yokoe2024Elbow,peng2023FunabotSuit}.
	This study aims to indirectly guide the hand position by inducing shoulder and elbow movements. Previous studies have succeeded in inducing elbow movements \cite{yokoe2024Elbow}; therefore, we focused on developing a fabric actuator that can provide McKibben-based surface tactile sensations to induce shoulder movements.
	
	\begin{figure}[!t]
		\centering
		\includegraphics[keepaspectratio=true,width=.65\linewidth]{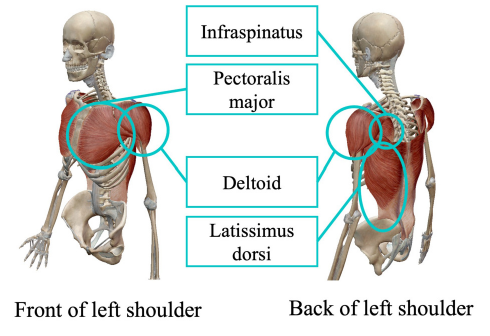}
		\caption{Muscles contributing to the shoulder movements \cite{visiblebody2024Human}.}
		\label{fig:muscles} 
	\end{figure}
	\begin{table}[!t]
		\caption{Shoulder Movements and Contributing Muscles.}
		\centering
		\begin{tabular}{c|c}
			Shoulder movement& Primary muscle contributing to movement\\\hline
			Abduction & Deltoid \\
			Adduction & Latissimus dorsi \\
			Horizontal abduction & Infraspinatus \\
			Horizontal adduction & Pectoralis major 
		\end{tabular}
		\label{tab:relation}
	\end{table}
	The proposed wearable fabric actuator consists of clothing (Motion Capture Suit, OptiTrack) and thin McKibben artificial muscles (EMM20 $\times$ 800 s-muscle). 
	The clothing is constructed with a Velcro loop on the exterior and the McKibben artificial muscles are attached to the clothing using Velcro tape. 
	When pneumatic pressure is applied to the artificial muscles, the clothing deforms, causing a McKibben-based tactile sensation.
	Accordingly, the arrangement of the McKibben artificial muscles within clothing is a crucial factor determining the movement induced by the McKibben-based surface tactile sensation. 
	Fig.~\ref{fig:muscles} illustrates the human muscles contributing to shoulder movements \cite{visiblebody2024Human}. 
	Table~\ref{tab:relation} lists the specific shoulder movements and the primary muscles contributing to these movements. 
	The proposed wearable fabric actuator provides tactile sensations to the body area around the four specific human muscles, as illustrated in Fig.~\ref{fig:muscles} to induce the four specific shoulder movements listed in Table~\ref{tab:relation}. 
	\begin{figure}[!t]
		\centering
		\includegraphics[keepaspectratio=true,width=.95\linewidth]{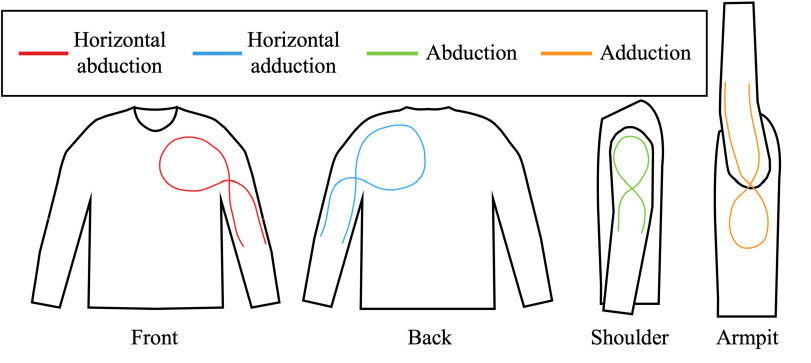}
		\caption{McKibben artificial muscle arrangement in the proposed fabric actuator. By applying pneumatic pressure to the artificial muscles, represented using the red, blue, green, and orange lines, the fabric actuator induces abduction, adduction, horizontal abduction, and horizontal adduction to the wearer's shoulder, respectively. }
		\label{fig:arrangement} 
	\end{figure}
	Based on Fig.~\ref{fig:muscles} and Table~\ref{tab:relation}, we arranged the McKibben artificial muscles of the proposed fabric actuator as shown in Fig.~\ref{fig:arrangement}. 
	This arrangement follows a similar strategy to our previous work~\cite{yokoe2024Elbow,peng2023FunabotSuit}. When the muscles contract in human movement, the fabric deforms in tandem with the muscle movement. Accordingly, the actuators are arranged to replicate deformation patterns similar to those of shoulder movement.
	This arrangement allows tactile sensations to be provided to the body areas near each human muscle, related to the four specific shoulder movements: abduction, adduction, horizontal abduction, and horizontal adduction. 
	
	\subsection{Evaluation of Eliciting Shoulder Movements by McKibben-based Surface Tactile Sensations}
	\label{sh}
	\subsubsection{Experimental setup}
\begin{figure}[!t]
		\centering
		\includegraphics[keepaspectratio=true,width=.7\linewidth]{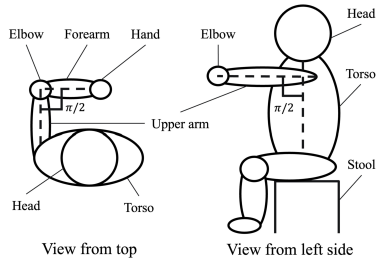}
		\caption{Schematic of participant's posture during the experiment. The elbow was bent at 90$^\circ$, and the left forearm was placed horizontally to the ground and in front of the body.}
		\label{fig:posture}
	\end{figure}

	We experimented with eight healthy participants to verify whether the wearer could identify the shoulder movements induced by the proposed wearable fabric actuator.
	The participants wore the proposed fabric actuator and a pair of earmuffs, sat on a stool, and assumed the posture illustrated in Fig.~\ref{fig:posture}.
	The fabric actuator randomly provided surface tactile sensations that induced four shoulder movements of the participant's left shoulder. The participants responded to the induced movements.
	The magnitude of the pneumatic pressure applied to the proposed fabric actuator was randomly determined from among seven values: 100, 150, 175, 200, 250, 300, and 350 kPa.
	All combinations of pneumatic pressure and induced movement were provided three times to each participant.
	Before the experiment, the participants were informed that they would perceive only one of the four movements. 
	However, were not informed regarding the tactile sensations corresponding to particular movements and whether their answers were correct.
	If the participant did not feel tactile sensation, participants refrained from responding.
	When participants responded, they were asked to report their confidence in their answers on a Likert scale from 1 to 7, where 1 indicated ``no confidence" and 7 indicated ``very confident." 
	Participants were not provided with any tactile sensations prior to the experiment to assess the intuitiveness of the proposed actuator.
	
	The performance of the proposed actuator was evaluated based on the number of correct responses. In addition, we calculated the precision, recall, and F1-score for each shoulder movement \cite{hand2021interpretable}. 
	
	\subsubsection{Experimental results}
	\label{sub:firstresult}
	\begin{table}[!t]
		\caption{Number of No, Correct, Wrong responses, and Participants' Confidence for each Pneumatic Pressure. }
		\centering
		\label{tab:confi}
		\begin{tabular}{c|ccc|c}
			\begin{tabular}[c]{@{}c@{}}Pneumatic\\ pressure {[}kPa{]}\end{tabular} &
			\begin{tabular}[c]{@{}c@{}}Without\\ responses\end{tabular} &
			\begin{tabular}[c]{@{}c@{}}Correct\\ responses\end{tabular} &
			\begin{tabular}[c]{@{}c@{}}Wrong\\ responses\end{tabular} &
			\begin{tabular}[c]{@{}c@{}}Mean of\\ confidence\end{tabular} \\ \hline
			100 & 83 & 10 & 2 & 1.1 \\
			150 & 28 & 60 & 4 & 1.6 \\
			175 & 6  & 82 & 4 & 1.9 \\
			200 & 1  & 88 & 3 & 3.0 \\
			250 & 1  & 90 & 3 & 4.2 \\
			300 & 0  & 92 & 1 & 5.0 \\
			350 & 0  & 89 & 2 & 5.4
		\end{tabular}
	\end{table}
	\begin{figure}[!t]
		\centering
		\includegraphics[keepaspectratio=true,width=.85\linewidth]{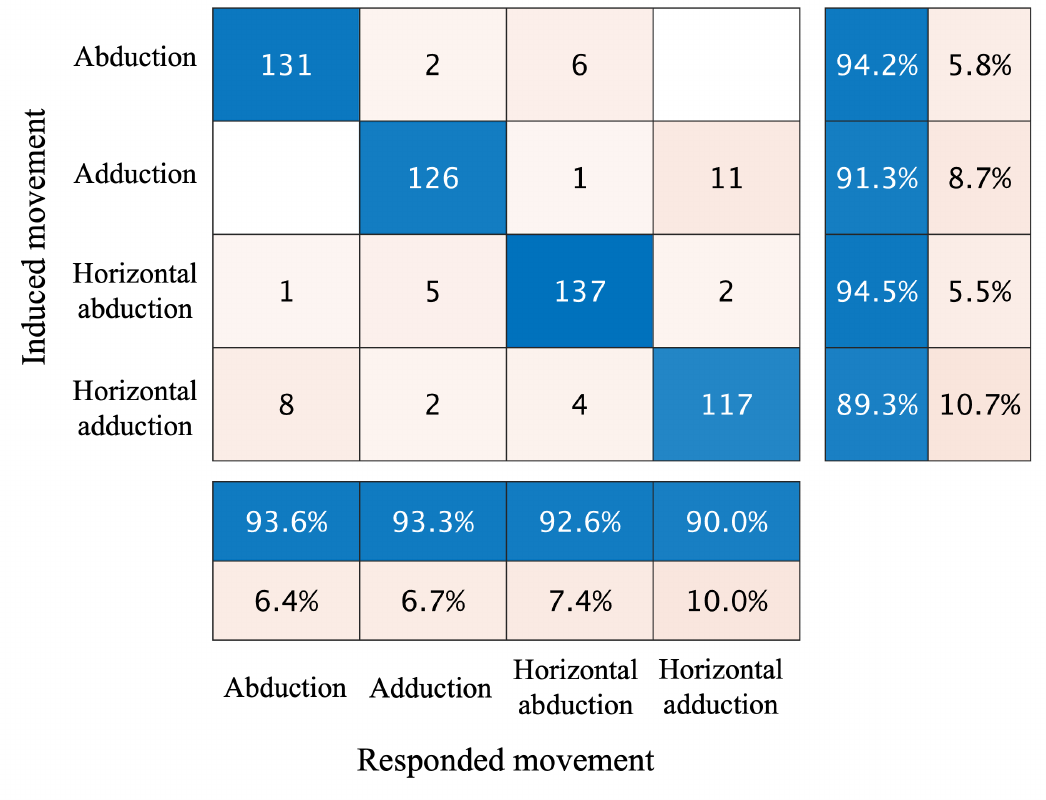}
		\caption{Confusion matrix of induced and responded movements. 
			The column and row represent induced and responded movements, respectively. The 2 $\times$ 4 matrix on the right shows the recall for each induced movement, and the 4 $\times$ 2 matrix on the bottom shows the precision for each responded movement.}
		\label{fig:confusion}
	\end{figure}
	\begin{table}[!t]
		\centering
		\caption{Precision, Recall, and F1-score for each Shoulder Movement.}
		\label{tab:rates}
		\begin{tabular}{c|ccc}
			Shoulder movement            & Precision & Recall & F1-score \\ \hline
			Abduction            & 0.94      & 0.94   & 0.94     \\
			Adduction            & 0.93      & 0.91   & 0.92     \\
			Horizontal abduction & 0.93      & 0.94   & 0.94     \\
			Horizontal adduction & 0.90      & 0.89   & 0.90    
		\end{tabular}
	\end{table}
	Table~\ref{tab:confi} lists the number of wrong, correct, no responses and participants' confidence in their responses for each pneumatic pressure. Confidence values represent the mean values measured when participants responded correctly.
		The $\chi^2$ test with Bonferroni correction revealed significant differences in correct and no responses between 100 kPa and all higher pressures, as well as between 150 kPa and all higher pressures (p$<$0.05). 
		As pressure increased from 100 kPa to 200 kPa, both correct responses and confidence ratings improved substantially, with diminishing effects at higher pressures (250--350 kPa). 
		This pattern suggests the relationship between pneumatic pressure and perceived tactile intensity follows the Weber-Fechner law \cite{colman2009Dictionary}.
	Fig.~\ref{fig:confusion} shows the confusion matrix of the induced movements and the movements responded to by the participants. 
	This confusion matrix only includes data when participants perceived tactile sensations and does not include data when they did not perceive tactile sensations.
	Table~\ref{tab:rates} lists the precision, recall, and F1-score for each shoulder movement. 
	The accuracy, considering all induced movements, is 0.92. The precision and recall of the shoulder movements are above 0.90 and 0.89, respectively.
	The F1-score, which considers the balance between precision and recall, also exceeds 0.90 for all shoulder movements.
	The results of the $\chi^2$ test with the Bonferroni correction showed no statistically significant differences in the number of correct responses per induced shoulder movement (p $>$ 0.05).
	The proposed fabric actuator can induce four shoulder movements without bias, with high precision and high recall. 
	These results validate our design strategy described in Section \ref{subsec:config}, as participants could intuitively perceive the tactile sensations without prior instruction. 
	
	\section{Hand Positional Guidance System Using the McKibben-based Surface Tactile Sensations}
	\subsection{Configuration of the Hand Positional Guidance System}
	\begin{figure}[!t]
		\centering
		\includegraphics[keepaspectratio=true,width=\linewidth]{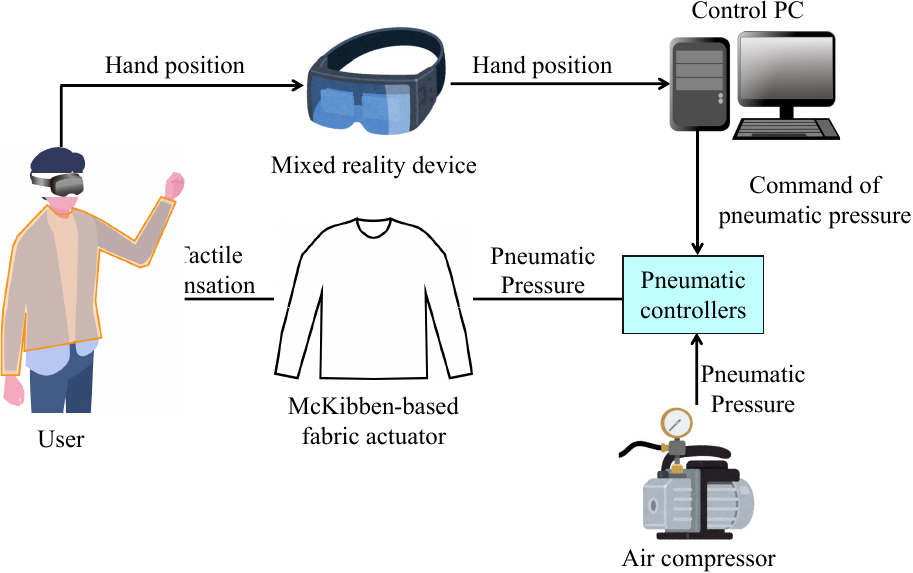}
		\caption{Configuration of the hand positional guidance system. The user wears the McKibben-based wearable fabric actuator and a mixed-reality device. Arrows indicate tactile presentation, communications, or pneumatic connections.}
		\label{fig:configuration}
	\end{figure}
	Fig.~\ref{fig:configuration} illustrates the configuration of the hand positional guidance system. 
	The proposed system comprises a control PC (Windows 10 Pro, 64-bit, CPU Intel (R) Core (TM) i7-11800H, 2.30 GHz, 32 GB RAM 3200 MHz, GPU NVIDIA GeForce RTX 3060 Laptop), a mixed-reality device with a hand tracking function (Hololens2, Microsoft), an air compressor (ACP-39SLB, TAKAGI), a pneumatic controller, and a wearable fabric actuator.
	The pneumatic controller comprises a programmable logic controller (007001001100, Industrial Shields) and electric, pneumatic regulators (CRCB-0135W/0136W, KOGANEI). 
	The wearable fabric actuator comprises clothing (Motion Capture Suit, OptiTrack) and six thin McKibben artificial muscles (EMM20 $\times$ 800, s-muscle). 
	The user of the proposed system wears the wearable fabric actuator and the mixed-reality device. 
	The mixed reality device tracks the hand position, which is transmitted to the control PC. The PC then calculates the commands for the pneumatic pressure and transmits the values to the pneumatic controller. Accordingly, the controller adjusts the pneumatic pressure to apply to the McKibben-based fabric actuator. The actuator provides McKibben-based surface tactile sensations to the user, who perceives these sensations and intuitively adjusts the hand position. 
	
	\begin{figure}[!t]
		\centering
		\includegraphics[keepaspectratio=true,width=.8\linewidth]{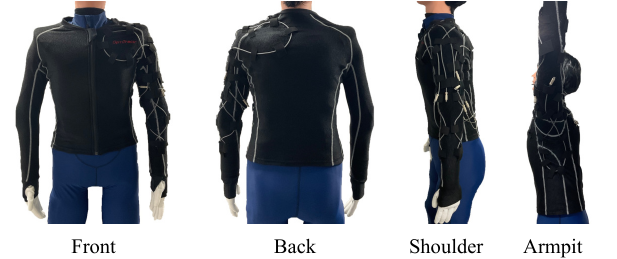}
		\caption{Overview of the fabric actuator used by hand positional guidance system. The wearable fabric actuator is dressed on a mannequin.}
		\label{fig:overview}
	\end{figure}
	Fig.~\ref{fig:overview} shows an overview of the fabric actuator used in the hand positional guidance system. 
	The artificial muscle arrangement of the actuator in the proposed system is constructed by combining the arrangement described in Section~\ref{sec:shoulder} and an arrangement capable of inducing flexion and extension of the elbow, as detailed in~\cite{yokoe2024Elbow}. 
	This configuration allows the proposed system to induce shoulder and elbow movements to indirectly guide the hand position.
	
	\subsection{Pneumatic Pressure Adjustment for Hand Positional Guidance}
	The proposed system indirectly  guides hand position by providing the user with six surface tactile sensations that induce shoulder and elbow movements. 
	\begin{figure}[!t]
		\centering
		\includegraphics[keepaspectratio=true,width=.48\linewidth]{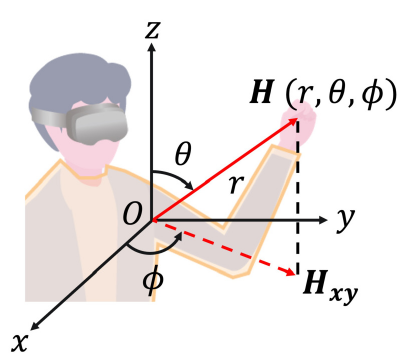}
		\caption{Modeling the hand position using a spherical coordinate system. The origin $O$ is the shoulder position, $x$, $y$, and $z$ directions denote the user's right direction, front, and upper directions, respectively. $\bm{H}$ represents the hand position vector. $\bm{H_{xy}}$ shows the vector of $\bm{H}$ projected on the $xy$-plane., $r$ shows the length of $\bm{H}$, $\theta$ represents the angle between the $z$-axis and $\bm{H}$, $\phi$ shows the angle between the $x$-axis and $\bm{H_{xy}}$. The direction of $r$, $\theta$, and $\phi$ are positive in the direction indicated by the arrows. }
		\label{fig:polor}
		\vskip -2mm
	\end{figure}
	\begin{table}[!t]
		\centering
		\caption{Relationship between Induced Movements and Parameter Adjustment Indicating Hand Positions.}
		\label{tab:relationship}
		\begin{tabular}{c|c|c}
			Parameters & Direction & Induced movement                \\ \hline
			$r$           & Positive  & Elbow extension               \\
			$r$           & Negative  & Elbow flexion                 \\
			$\theta$      & Positive  & Shoulder adduction            \\
			$\theta$      & Negative  & Shoulder abduction            \\
			$\phi$        & Positive  & Shoulder horizontal abduction \\
			$\phi$        & Negative  & Shoulder horizontal adduction
		\end{tabular}
	\end{table}
	Fig.~\ref{fig:polor} shows the position vector of the hand in the spherical coordinate system. 
	The proposed system adjusts parameters $r$, $\theta$, and $\phi$, shown in Fig.~\ref{fig:polor}, by inducing shoulder and elbow movements to guide the hand position indirectly.
	Table~\ref{tab:relationship} lists the movements induced by the proposed system when the system adjusts the hand parameters to positive or negative values.
	
	The intensity of the McKibben-based surface tactile sensations varies with the magnitude of the pneumatic pressure applied to the fabric actuator.
	Consequently, the proposed system necessitates an adjustment algorithm for the pneumatic pressure applied to the actuator, contingent upon the magnitude of parameter adjustment.
	The proposed system implements an adjustment algorithm for pneumatic pressure based on the following four assumptions: 
	\begin{enumerate}
		\item Parameters $r$, $\theta$, and $\phi$ are independent.
		\item The induced movements to adjust $r$, $\theta$, and $\phi$ follow the equilibrium point hypothesis \cite{feldman1986Once}. 
		\item Each equilibrium point varies linearly with the human's perception of the McKibben-based surface tactile sensation.
		\item The human perception of surface tactile sensation and magnitude of pneumatic pressure applied to the fabric actuator follow the Weber--Fechner law \cite{colman2009Dictionary}.
	\end{enumerate}
	Based on these assumptions, we can implement traditional PID control while handling the nonlinear characteristics between pneumatic pressure and tactile perception demonstrated in Section~\ref{sub:firstresult}.
	Define vector $\bm{q}$ corresponding to elbow extension, elbow flexion, shoulder adduction, shoulder abduction, shoulder horizontal abduction, and shoulder horizontal adduction when guiding the hand position as follows:
	\begin{align}\label{eq:q}
		\bm{q}=\left[ r,-r,\theta,-\theta,\phi,-\phi \right]^\mathsf{T}.
	\end{align}
	We used six parameters to represent both positive and negative directions of movement, ensuring a one-to-one mapping between these parameters and tactile sensations.
	According to the equilibrium point hypothesis, humans move their bodies by shifting their equilibrium points. 
	Therefore, the equilibrium point of each parameter corresponding to the target hand position must be determined.
	The equilibrium point vector $\bm{q_{eq}}$ represents the required parameter shifts for the target hand position. To account for the nonlinear relationship between pneumatic pressure and tactile perception, we define $\bm{q_{eq}}$ as follows:
	\begin{multline}
		\label{eq:pid}
		\bm{q_{eq}}=K_P\left(\bm{q_{eq}^d}-\bm{q}\right)
		+K_D\frac{d}{dt}\left(\bm{q_{eq}^d}-\bm{q}\right)\\
		+K_I\int \left(\bm{q_{eq}^d}-\bm{q}\right) dt,
	\end{multline}
	where $\bm{q_{eq}^d}$ is the target equilibrium point vector and $K_P$, $K_D$, and $K_I$ are constants.
	Based on the Weber--Fechner law, the pneumatic pressure $p_i~(i=1,\cdots, 6)$ applied to the McKibben artificial muscles to induce elbow and shoulder movements is as follows: 
	\begin{align}\label{eq:air}
		p_i=p^0_i\exp\left({\frac{q_{eq, i}}{k_i}}\right) \hskip 2mm (i=1,\cdots, 6),
	\end{align}
	where $q_{eq, i}$ is the $i$-th component of $\bm{q_{eq}}$, $p^0_i$ is the minimum threshold of pneumatic pressure applied to each artificial muscle for the user to perceive the surface tactile sensation, and $k_i$ is a constant corresponding to each artificial muscle. 
	This two-stage control approach enables intuitive tactile guidance by integrating traditional control theory (Equation~\eqref{eq:pid}) with human sensory perception (Equation~\eqref{eq:air}), mapping equilibrium shifts to pneumatic pressures while accounting for their nonlinear perceptual relationship.
	The minimum thresholds $p^0_i~(i=1,\cdots, 6)$ are measured individually for each user and body region where tactile sensations are provided using the method of limits \cite{colman2009Dictionary} to account for variations in tactile sensitivity. 
	This study determines $k_i$ as follows: 
	\begin{align}\label{eq:k}
		k_i=\frac{q_{eq, i}^M}{\ln\left(\frac{p^M_i}{p^0_i}\right)} \hskip 2mm (i=1,\cdots, 6),
	\end{align}
	where $q_{eq, i}^M$ and $p^M_i$ are the maximum value of $q_{eq, i}$ and $p_i$, respectively. 
	From the definition of $k_i$, the proposed system applies the maximum pneumatic pressure $p^M_i$ to an artificial muscle to guide the maximum equilibrium point $q_{eq, i}^M$. 
	The maximum equilibrium point vector $\bm{q_{eq}^M}$ can be adjusted based on application requirements. While increasing $\bm{q_{eq}^M}$ enables guidance across the full anatomical range, it reduces tactile sensation gradients and may compromise guidance performance.
	
	\section{Evaluation of Hand Positional Guidance System}
	\subsection{Accuracy Evaluation of Hand Positional Guidance}
	\label{ac}
	\subsubsection{Experimental setup}
	We experimentally evaluated the accuracy of the proposed hand positional guidance system.
	\begin{figure}[!t]
		\centering
		\includegraphics[keepaspectratio=true,width=.95\linewidth]{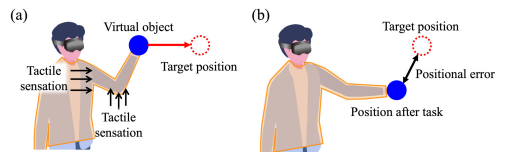}
		\caption{Accuracy evaluation procedure of hand positional guidance system. (a) Participant manipulates a virtual object to the target position following the guidance. (b) System measures the positional error between the target position and virtual object after an experimental task. }
		\label{fig:procedure_ac}
	\end{figure}
	\begin{figure}[!t]
		\centering
		\includegraphics[keepaspectratio=true,width=.5\linewidth]{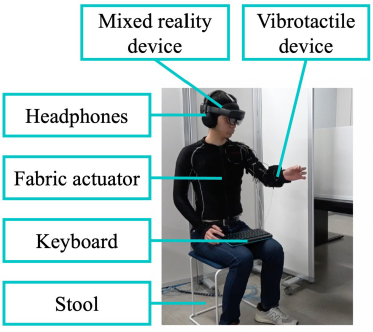}
		\caption{Overview of the participants during the experiment. }
		\label{fig:subject}
	\end{figure}
	Fig.~\ref{fig:procedure_ac} illustrates the procedure to evaluate the accuracy of the hand positional guidance system. 
	Fig.~\ref{fig:subject} presents an overview of the participants during the experiment. 
	The participants sat on a stool, placed a keyboard on their lap, and wore the proposed fabric actuator, a vibrotactile device (Tactosy for Arms, bHaptics), a mixed-reality device (Hololens2, Microsoft), and a set of headphones. 
	The participants were instructed to move a virtual object according to the guidance using their left hand.  
	The participants pressed the keyboard when they finished manipulating the virtual object. The accuracy of the hand positional guidance system was evaluated by measuring the error between the position of the virtual object and target position.
	The virtual object was a sphere of 50 mm diameter, located at $r$ = 400 mm, 
	$\theta$ = 90$^\circ$, and $\phi$ = 90$^\circ$ in the coordinate system shown in Fig.~\ref{fig:polor}.
	Additionally, the target position was randomly determined for each task in the range of 200--600 mm for $r$ and 
	40--140$^\circ$ for $\theta$ and $\phi$.
	The maximum equilibrium point vector $q_{eq, i}^M (i=1,\cdots, 6)$ was set to 600 mm, -200 mm, 140$^\circ$, -40$^\circ$, 140$^\circ$, and -40$^\circ$, respectively.
	
	Each participant performed this task under three conditions: 
	In Condition 1, the proposed system, that is, the McKibben-based surface tactile sensation, was used to guide the hand position.
	Prior to the experiment, the participants were informed that a tactile sensation would guide their hand position but not the nature of the tactile sensation.
	In Condition 2, the participants were made to listen to the synthesized speech through headphones and adjusted the position of their hands accordingly. 
	Speech provided values of $r$, $\theta$, and $\phi$, as shown in Fig.~\ref{fig:polor}, for the target position only once at the beginning of each task. The resolution of the speech output was 10 mm for $r$ and 1$^\circ$ for $\theta$ and $\phi$.
	Prior to the experiment, the participants were informed how the synthesized speech would be output.
	In Condition 3, vibrotactile guided the participants' hand positions. 
	The vibrotactile presentation method was based on a previous study using ``Phantom Sensation" \cite{salazarluces2018PhantomSensation}. 
	However, this previous study provided only hand positional guidance on the $xz$-plane in Fig.~\ref{fig:polor}.
	Therefore, we used a vibrotactile device with vibrators, with half the device on the wrist side and half on the elbow-side. 
	The wrist- and elbow-side vibrators were used when guiding the hand to positive and negative values in the $y$-axis in Fig.~\ref{fig:polor}, respectively. 
	Prior to the experiment, the participants were informed which vibrotactile sensation would guide their hand to which position. 
	From the perspective of the leading joint hypothesis discussed in Section~\ref{sec:related}, the vibrotactile guidance system uses the wrist as the leading joint, while our McKibben-based system uses the shoulder and elbow as leading joints. This comparison allows us to evaluate which is more suitable for the leading joint for hand position guidance.
	No prior practice was conducted to familiarize the participants with the guidance under all conditions.
	Nine healthy individuals participated in this experiment. 
	All participants performed the experimental task ten times under each condition, with the order of the conditions randomly counterbalanced.
	
	\subsubsection{Experimental result}
	\begin{figure*}[!t]
		\centering
		\includegraphics[keepaspectratio=true,width=.8\linewidth]{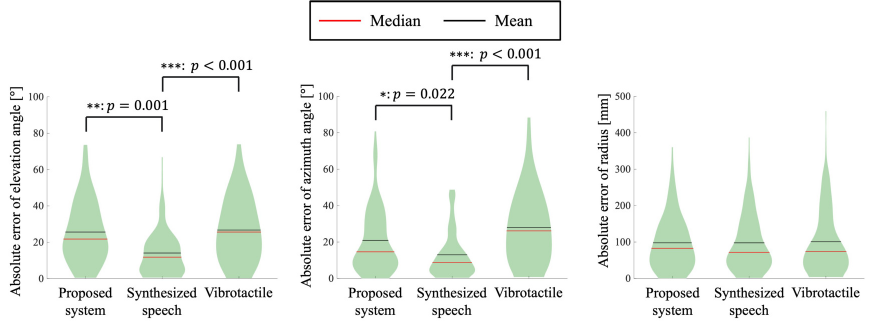}
		\caption{Violin plots of the absolute positional error between the target position and virtual object after an experimental task. $^*$, $^{**}$, and $^{***}$ represent statistically significant differences based on the Friedman test with Bonferroni correction \cite{dunn1964Multiple}. }
		\label{fig:violin_ac}
	\end{figure*}
	\begin{table}[!t]
		\centering
		\caption{Median values of the absolute error between the target position and virtual object manipulated by the participants.}
		\label{tab:median}
		\begin{tabular}{c|ccc}
			Guidance method &
			\begin{tabular}[c]{@{}c@{}}Elevation\\ angle {[}deg{]}\end{tabular} &
			\begin{tabular}[c]{@{}c@{}}Azimuth\\ angle {[}deg{]}\end{tabular} &
			Radius {[}mm{]} \\ \hline
			Proposed system    & 21.8 & 14.6 & 82.8 \\
			Synthesized speech & 11.8 & 8.8  & 71.8 \\
			Vibrotactile       & 25.6 & 26.1 & 73.9
		\end{tabular}
	\end{table}
	Fig.~\ref{fig:violin_ac} shows the violin plots of the absolute error between the target position and virtual object manipulated by the participants.
	Table~\ref{tab:median} lists the median values of the absolute error between the target position and  virtual object for each guidance method. 
	Statistically significant differences are observed between the proposed system and synthesized speech in the absolute errors of the elevation and azimuth angles ($p$ = 0.001 and 0.022: Friedman test with Bonferroni correction). 
	In contrast, there are no statistically significant differences between the proposed system and vibrotactile for elevation and azimuth angles; however, the median values of the proposed system are smaller than those of the vibrotactile.
	In addition, statistically significant differences are observed between the vibrotactile and synthesized speech in the absolute errors of elevation and azimuth angles ($p$ $<$ 0.001 and 0.001: Friedman test with Bonferroni correction). 
	However, there are no statistically significant differences in the absolute errors of radius. 
	The results demonstrate that the proposed system can guide hand position with the same accuracy as the vibrotactile sensation without prior instruction.
	
	\subsection{Speed Evaluation of Hand Positional Guidance}
	\label{sp}
	\subsubsection{Experimental setup}
	Additionally, we assessed the speed at which the proposed system guides the hand position through an experiment with participants.
	\begin{figure}[!t]
		\centering
		\includegraphics[keepaspectratio=true,width=.9\linewidth]{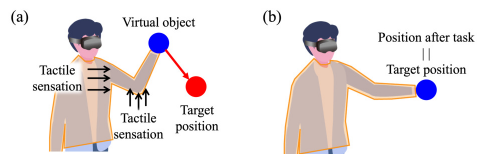}
		\caption{Speed evaluation procedure of hand positional guidance system. (a) Participant manipulates the blue virtual object to the red target virtual object following the guidance. (b) The participant is required to move the virtual object to the target object repeatedly. }
		\label{fig:procedure_sp}
	\end{figure}
	Fig.~\ref{fig:procedure_sp} shows the procedure for speed evaluation of the hand positional guidance system. 
	As in Section~\ref{ac}, the participants in this experiment sat on a stool and wore the proposed fabric actuator, a vibrotactile device, a mixed-reality device, and a set of headphones. 
	\begin{figure}[!t]
		\centering
		\includegraphics[keepaspectratio=true,width=.55\linewidth]{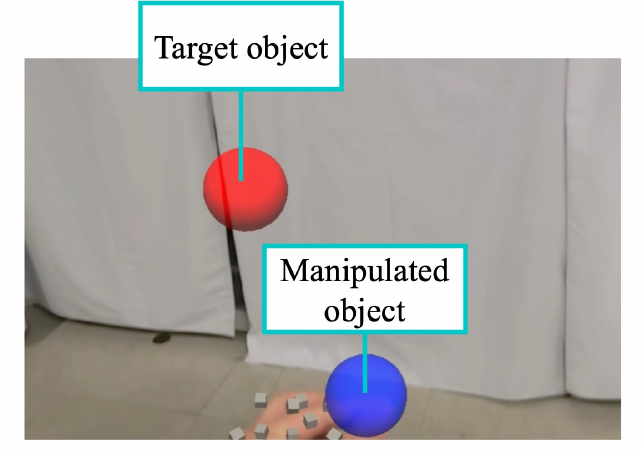}
		\vskip -2mm
		\caption{Immersive environment as seen by the participant during the experiment. The participant manipulates the blue virtual object with their hand so that it touches the red virtual object. The virtual objects are spheres of 50 mm diameter. }
		\label{fig:immersive}
	\end{figure}
	Fig.~\ref{fig:immersive} depicts the immersive environment observed by the participants during the experiment.
	The participants manipulate the virtual object in blue with their left hand for 40 s following the guidance to touch the target virtual object in red, which appear at a random position. 
	At the outset of the experiment, the blue virtual object was located at $r$ = 400 mm, $\theta$ = 90$^\circ$, and $\phi$ = 90$^\circ$ in the coordinate system as shown in Fig.~\ref{fig:polor}. 
	The position of the red virtual object was randomly determined within the ranges of 200--600 mm for $r$ and 40--140$^\circ$ for $\theta$ and $\phi$. 
	When the blue and red virtual objects touched, the red virtual object moved to $r$ = 400 mm, $\theta$ = 90$^\circ$, and $\phi$ = 90$^\circ$, that is, the initial position of the blue virtual object. 
	Once the participant returned the blue virtual object to its initial position, the red virtual object was moved to a random position, and the process was repeated.
	The maximum equilibrium point vector $q_{eq, i}^M (i=1,\cdots, 6)$ was set to same values as in Section \ref{ac}.
	
	The speed of hand positional guidance was evaluated based on the number of red virtual objects placed at random positions could be touched in 40 s by repeating the manipulation of the blue virtual object. 
	
	The participants experimented under three conditions, each with a different hand positional guidance method introduced in Section \ref{ac}. Nine healthy participants performed the experiments five times per condition.
	No prior practice was conducted to familiarize the participants with hand positional guidance.
	
	\subsubsection{Experimental result}
	\begin{figure}[!t]
		\centering
		\includegraphics[keepaspectratio=true,width=.55\linewidth]{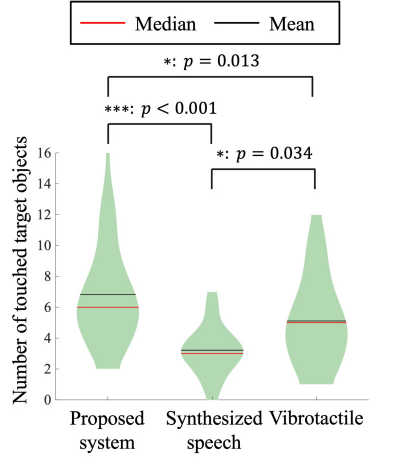}
		\caption{Violin plot of the number of red virtual objects touched with the blue virtual object.  $^*$ and $^{***}$ represent statistically significant differences based on Friedman test with Bonferroni correction \cite{dunn1964Multiple}. }
		\label{fig:violin_sp}
	\end{figure}
	Fig.~\ref{fig:violin_sp} shows a violin plot representing the number of red virtual objects touched by the blue virtual object in 40 s. 
	The proposed system achieves the highest median and mean scores for the three conditions.
	In addition, the results of the Friedman test with Bonferroni correction yield statistically significant differences among all the conditions.
	These results indicate that the proposed system can guide the hand position at the highest speed among the three conditions without prior instruction.
	
	\section{Discussion}
	As mentioned in Section~\ref{sh}, the evaluation of shoulder
	movement elicitation using the proposed wearable fabric actuator demonstrated that the actuator could induce all four movements (abduction, adduction, horizontal abduction, and horizontal adduction) with high accuracy (92\%). 
	Additionally, the high F1-scores ($>$ 0.9) for each shoulder movement and the lack of a statistically significant difference among all four movements indicate that the actuator can induce shoulder movements through the McKibben-based surface tactile sensation without bias toward any movement.
	These results suggest that the arrangement of the McKibben artificial muscles illustrated in Fig.~\ref{fig:arrangement} effectively provides tactile sensations to the skin around the muscles that contribute to shoulder movements, allowing intuitive tactile perception.
	Moreover, the participants could intuitively perceive the movements based on the deformation of clothing, which participants wear daily, by the McKibben artificial muscles.
	
	The experiment described in Section~\ref{ac} revealed that hand positional guidance using synthesized speech was more accurate than that using tactile sensations for azimuth and elevation angles, as shown in Fig.~\ref{fig:polor}. 
	In the experiment, the speech resolution was 1$^\circ$, which may exceed the resolution of human tactile perception.
	However, the experiment described in Section~\ref{sp} revealed that the speed of hand positional guidance using synthesized speech was slower than that using tactile sensations.
	Synthesized speech requires time to listen to and understand speech information because of the limited amount of output per unit of time. These times may have resulted in slower hand positional guidance than that of tactile sensation. 
	It is recommended to use tactile information for time-constrained tasks and speech for tasks requiring high accuracy but not time-constrained.
	
	In our experiment, the participants were not informed of the relationship between the target hand position and tactile sensation when using the proposed system. 
	Nevertheless, the proposed system could guide the hand position with the same accuracy as that of the vibrotactile, where participants need to be informed of the relationship between the target hand position and vibrotactile.
	In addition, the proposed system has a higher hand guidance speed than that of vibrotactile.
	The results suggest that the proposed system is highly intuitive, allowing users to effectively perceive tactile sensations as well as realize accurate and fast hand positional guidance without prior instruction on the relationship between the target position and tactile sensation.
	
	\section{Conclusion}
	This paper presents a hand positional guidance system utilizing McKibben-based surface tactile sensations provided to the shoulder and elbow. 
	We developed a wearable fabric actuator to provide the tactile sensations that induce shoulder movements. 
	Our actuator demonstrated a high accuracy in inducing four specific shoulder movements without bias, confirming the effectiveness of the McKibben artificial muscle arrangement. 
	The proposed system induces shoulder and elbow movements by combining the proposed McKibben artificial muscle arrangement with the artificial muscle arrangement of the elbow presented in a previous study.
	The proposed system uses a psychophysics-based algorithm for tactile sensation, allowing users to intuitively adjust their hand position based on the McKibben-based surface tactile sensation that induces shoulder and elbow movements. 
	
	Experiments and statistical analyses demonstrated a trade-off between the accuracy and speed of hand positional guidance when comparing the proposed system with that using synthesized speech. Our system exhibited higher speed in hand-positioning tasks, whereas speech provided higher accuracy guidance. Notably, our system achieved comparable accuracy and a higher speed than that of vibrotactile guidance, even without prior instructions on tactile perception, highlighting the intuitiveness of the proposed system. 
	The results of this study demonstrate the potential for improving the user experience in immersive environments through intuitive, instruction-free hand positional guidance, which could lead to more efficient and user-friendly interactions in various applications such as virtual reality, teleoperation, and rehabilitation.
	Future work will include hand positional guidance for active target objects moving in immersive environments and application to motor learning, as in \cite{matsui2024Realtime}.
	
	\bibliography{ral_handposition}
	\bibliographystyle{IEEEtran}
	
	\vfill
\end{document}